\documentclass[twocolumn,preprintnumbers,showpacs]{revtex4}
\usepackage{amsmath}
\usepackage{dcolumn}
\usepackage{bm}
\usepackage{graphicx}
\usepackage{amsfonts}
\usepackage{amssymb}
\usepackage{mathrsfs}
\usepackage{color}
\usepackage{braket}

\begin{document}

\title{Quantum speed-up in solving the maximal clique problem}
\author{Weng-Long Chang$^{1}$}
\email{changwl@cc.kuas.edu.tw}
\author{Qi Yu$^{2}$}
\author{Zhaokai Li$^{2,3}$}
\author{Jiahui Chen$^{2,4}$}
\author{Xinhua Peng$^{2,3,6}$}
\email{xhpeng@ustc.edu.cn}
\author{Mang Feng$^{5,6,7}$}
\email{mangfeng@wipm.ac.cn}
\affiliation{$^{1}$ Department of Computer Science, National Kaohsiung University of Applied Sciences, Kaohsiung City 80778, Taiwan, China \\
$^{2}$ CAS Key Laboratory of Microscale Magnetic Resonance and Department of Modern Physics, University of Science and Technology of China, Hefei 230026, China \\
$^{3}$ Synergetic Innovation Center of Quantum Information and Quantum Physics, University of Science and Technology of China, Hefei, China \\
$^{4}$ Institute for Quantum Computing and Department of Physics and Astronomy, University of Waterloo, Waterloo, Ontario, Canada \\
$^{5}$ State Key Laboratory of Magnetic Resonance and Atomic and Molecular Physics, Wuhan Institute of Physics and Mathematics, Chinese Academy of Sciences, Wuhan, 430071, China\\
$^{6}$ Synergetic Innovation Center for Quantum Effects and Applications (SICQEA), Hunan Normal University, Changsha, 410081, China \\
$^{7}$ Department of Physics, Zhejiang Normal University, Jinhua, 321004, China }

\begin{abstract}
The maximal clique problem, to find the maximally sized clique in a given graph, is classically an NP-complete computational problem, which has potential applications ranging from  electrical engineering, computational chemistry, bioinformatics to social networks. Here we develop a quantum algorithm
to solve the maximal clique problem for any graph $G$ with $n$ vertices with quadratic speed-up over its classical counterparts, where the time and spatial complexities are reduced to, respectively, $O(\sqrt{2^{n}})$ and $O(n^{2})$. With respect to oracle-related quantum algorithms for the NP-complete problems, we identify our algorithm to be optimal.
To justify the feasibility of the proposed quantum algorithm, we have successfully solved an exemplified clique problem for a graph $G$ with two vertices and one edge by carrying out a nuclear magnetic resonance experiment involving four qubits.
\end{abstract}
\maketitle

\section{introduction}

In a social network, identifying the largest group of people with mutual acquaintance, i.e., who all know each other, is an NP-complete problem \cite{clique}, whose complexity scales exponentially with the number of the persons involved in the social network. It is mathematically termed the maximal clique problem \cite{karp,garey}, described by a graph with vertices and edges representing, respectively, the persons and their mutual relations. Its solution is to find the largest group (or groups) with the most vertices connected mutually by edges. Besides its applications in social networks, the clique problem has also been applied to electrical engineering for designing efficient circuits \cite{circuit}, computational chemistry for exploring bound chemicals in the database \cite{database} and bioinformatics for studying evolutionary trees of species or predicting protein structure \cite{info}.

\begin{figure}
\centering {\includegraphics[width=7.2 cm, clip=True]{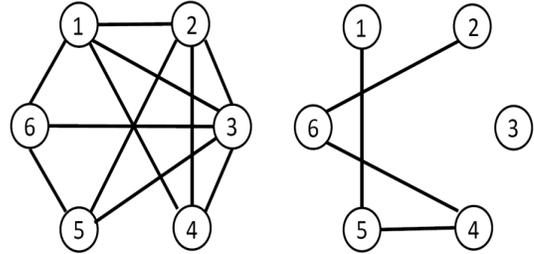}}
\caption{An exemplified graph with six vertices and eleven edges (Left Panel) as well as its complementary graph (Right Panel). As mentioned in the text, using the complementary graph, we may find the illegal cliques more easily by identifying the vertices connected by edges.}
\label{fig1}
\end{figure}

Mathematically, the maximal clique problem is defined regarding a graph $G=(V,E)$ with $n$ vertices and $\theta$ edges, where $V$ is a finite set of the $n$ vertices in $G$ and $E$ is a set of the $\theta$ edges connecting pairs of vertices in $G$. A clique is a set of vertices in which all the vertices are connected with each other by edges.  As such, the maximal clique problem is to find the largest clique in the graph, which has been proven to be NP-complete. Fig. \ref{fig1} shows an example of a graph with six vertices for such a problem where the vertices $\{1,2,3,4\}$ form the largest clique. It has been shown that any brute-force solution to the maximal clique problem requires an exponential increase of time with the size of the problem (i.e., with time complexity of $O(2^{n})$) \cite{karp,garey} and no effective approximation is found to solve the clique problem \cite{feige}. DNA computing techniques, under the condition of exponentially increasing volumes of DNA (spatial complexity), claimed to solve this problem in linearly increasing time due to operations in parallel \cite{Science1997,chang-dna}. But this is not true in real operations since the maximum number of vertices in procession is limited to 27 \cite{Science1997}.

Quantum computers promise to exploit the remarkable properties of quantum mechanical systems to solve certain problems more efficiently than their classical counterparts. Besides the celebrated Shor's algorithm for integer factorization \cite{shor} and Grover's algorithm for searching unsorted database \cite{grover}, some other quantum algorithms for various hard problems have recently been proposed \cite {linear,bs,field,squarefree}. On the other hand, experimental progress has witnessed the successful implementation of various quantum algorithms, such as factoring algorithm in different quantum computer candidates \cite{shor1,shor2,shor3,shor4,fact1,shor5,shor6} and efficient execution of boson sampling \cite{bosonsamping1,bosonsamping2,bosonsamping3,bosonsamping4}. However, some of the aforementioned work are not for NP-complete problems. In principle, the NP-complete problems can be solved by the oracle-related search algorithms, such as Grover search. It is already known that any of such oracle-based quantum algorithms could not perform better than quadratic speed-up over its classical counterparts \cite{NP1,NP2}. This implies that an oracle-related quantum algorithm for the maximal clique problem, if behaving optimally with finite spatial complexity, should work in time $O(\sqrt{2^{n}})$.

Quantum adiabatic algorithm has been proposed to solve the maximal clique problem \cite{adia}. Unfortunately, asymptotic analysis of quantum adiabatic evolution algorithms appears to be difficult.
Here we propose an optimal oracle-related quantum algorithm, based on a quantum circuit oracle model, for solving the maximal clique problem with quadratic speed-up over its classical counterparts. By representing the vertices by qubits, we first filter out the illegal cliques (defined later) under quantum logical gates; then we identify the maximally sized subset of vertices in the legal cliques, followed by operations of Grover search for the target states representing the solutions of the clique problem. The key point of our algorithm is the polynomial time complexity of the oracle's job for labeling the target states. As a justification of the feasibility of our algorithm, a four-qubit nuclear magnetic resonance (NMR) experiment is accomplished to solve an exemplified clique problem for a graph $G$ with two vertices and one edge.

\section{the quantum algorithm}

For a graph with $n$ vertices, we require $n$ qubits representing the vertices, and there are 2$^{n}$ possible cliques from $|0\cdots 0\rangle$ to $|1\cdots 1\rangle$, where $|0\cdots 0\rangle$ and $|1\cdots 1\rangle$ represent, respectively, the clique with no vertex and the one with $n$ vertices. The qubit state $|1\rangle$ ($|0\rangle$) represents the presence (absence) of the corresponding vertex in the clique. For the example in Fig. 1, the maximally sized set $\{1,2,3,4\}$ can be represented as $|111100\rangle$. An efficient way to solve a clique problem of the graph is to consider its complementary graph $\bar{G}=(V,\bar{E})$ with the edges of the vertices out of the set $E$.
For convenience of description, we introduce the definitions of legal clique and illegal clique. A legal clique is the one with no edge in $\bar{G}$ and thus other cliques in $\bar{G}$ belong to illegal cliques (i.e., the vertices connected by edges in $\bar{G}$ form illegal cliques). After excluding the illegal cliques by means of $\bar{G}$, we are able to find the solution from the legal cliques by identifying the maximally sized subset of vertices in $G$.
For example, the state $|1x_2 x_3x_41x_6\rangle$ with $x_i=$1 or 0 for the graph in Fig. 1, evidently denotes an illegal clique. After removing the illegal cliques, we then explore the maximally sized set from the legal cliques as the solution and set it as a target state. The final step is to find the target state by iterating the Grover search operations.

\subsection{The steps of the solution}

\begin{figure*}[t]
\centering {\includegraphics[width=15.2 cm, clip=True]{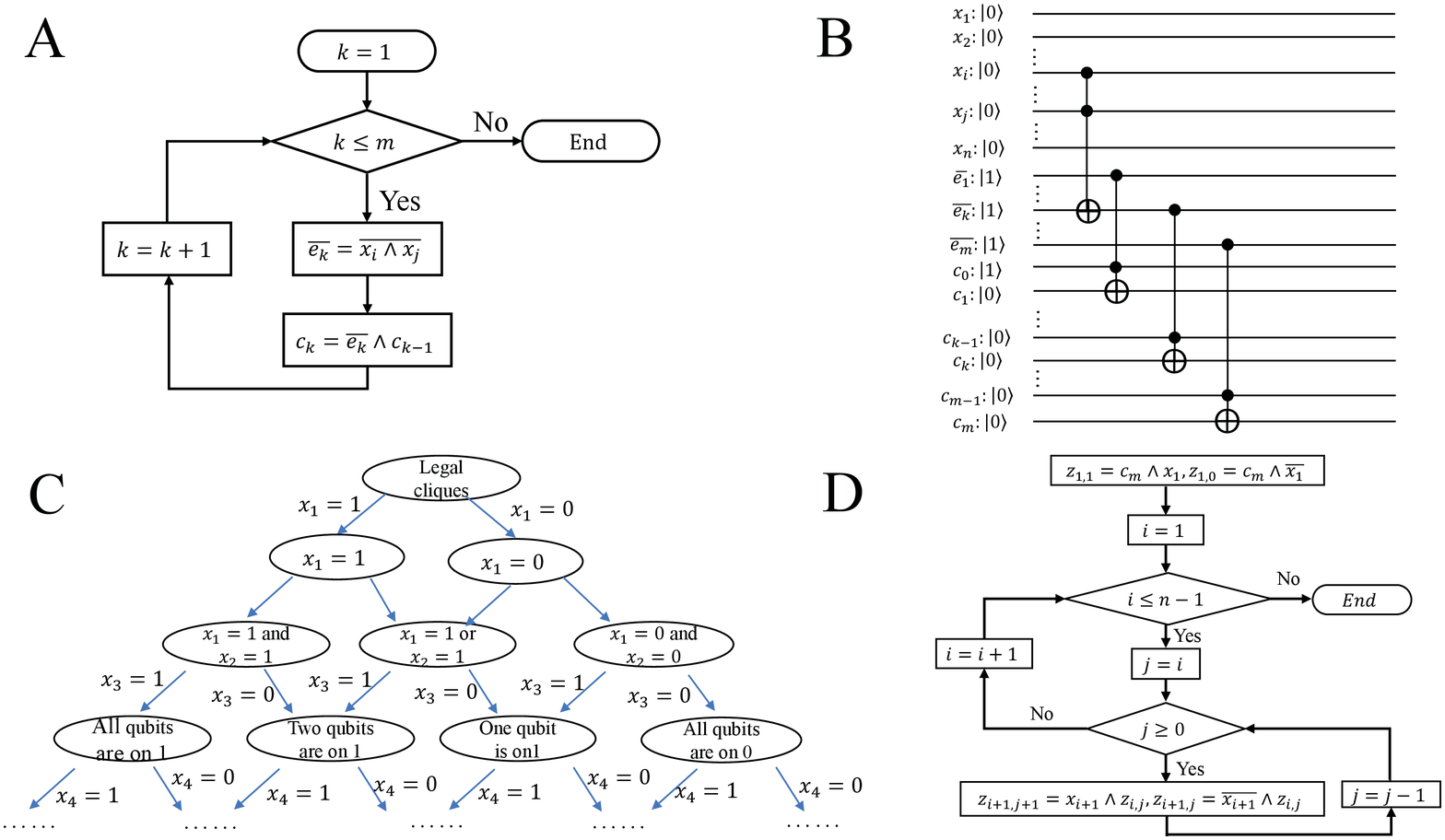}}
\caption{Quantum mechanical treatment of the maximal clique problem of a graph. ({\bf A}) Logical flowchart of deciding a legal clique. The formula $\left( {\overline {{x_i} \wedge {x_j}} } \right)$ is realized by the Toffoli gate when the target qubit is in $|1\rangle$ and the formula $({\bar{e}_k}\wedge {c_{k - 1}})$ is achieved by the Toffoli gate when the target qubit is in $|0\rangle$. Following the steps, we continue the loop by increasing $k$. ({\bf B}) Quantum circuit corresponding to ({\bf A}). $x_{k}$ ($k=1,2,\cdots$) represent states of qubits encoding the $k$th vertex $v_k$. $\bar{e}_{k}$ ($k=0,1,2,\cdots$) mean states of auxiliary qubits for storing states of the illegal and legal cliques, which are all initialized as $|1\rangle$. $c_{k}$ ($k=0,1,2,\cdots$) are states of auxiliary qubits for further assistance as explained in the flowchart. Here, $c_0$ is initialized as $|1\rangle$ and the rest are initially set to be $|0\rangle$. ({\bf C}) Tree diagram for classifying the legal cliques according to the Hamming weight of $x$.  In the diagram, we first divide the legal cliques into two registers conditional on the value of $x_1$. Then we divide the two registers into three depending on the value of $x_2$. After $n$ steps, we can classify all the legal cliques into $(n+1)$ registers. These registers contain the cliques with different numbers of vertices from 0 to $n$. ({\bf D}) Logical flowchart for counting the number of the vertices. The formulae $\left( {{x_{i + 1}} \wedge {z_{i,j}}} \right)$ and $\left( {\overline {{x_{i + 1}}}  \wedge {z_{i,j}}} \right)$ are carried out cyclically to store their  results in the auxiliary qubits $z_{i+1,j+1}$ and $z_{i+1,j}$, respectively. Implementation of $\left( {{c_m} \wedge {x_1}} \right)$ and $\left( {{c_m} \wedge \overline {{x_1}}}\right)$ in the auxiliary qubits $z_{1,1}$ and $z_{1,0}$ is the beginning of the loop. }
\label{fig2}
\end{figure*}

The concrete implementation of the quantum algorithm is as follows:

1) {\textbf{Preparing a uniform superposition state.}} We first prepare a uniform superposition state of $n$ qubits $ \frac{1} {\sqrt{2^n}} \sum_{x = 0}^{2^{n}-1} \vert x \rangle$ involving all $2^n$ possible cliques by individually performing Hadamard gate on each qubit initially prepared in $|0\rangle$. We call this register as the data register.

2) {\textbf{Excluding the illegal cliques.}} For the graph $G$ with $n$ vertices and $\theta$ edges, it is easy to find its complementary graph $\bar{G}$ with $n$ vertices and $m=n(n-1)/2-\theta$ edges.  Any two vertices $v_{i}$ and $v_{j}$ disconnecting in the original graph $G$ are connected in the complementary graph $\bar{G}$, i.e., the edges in $\bar{G}$ are represented by $\bar{e}_k=(v_i, v_j)$, where $1\leq k \leq m$. Therefore we remove the sets represented by $|x_1 x_2 \cdots 1_{i}\cdots 1_{j}\cdots x_{n-1} x_n\rangle$ from all possible cliques. If the $k$th edge exists in  $\bar{G}$, the formula $\left({\overline {{x_i} \wedge {x_j}} } \right)$ is then of the true value. As such, the requested condition for deciding a legal clique among $2^n$ possible cliques is that the formula $\wedge_{k = 1}^m\left( {\overline {{x_i} \wedge {x_j}} } \right)$ is true. To accomplish the logical flowchart in Fig. \ref{fig2}A in quantum computer, we have to introduce some auxiliary qubits, such as $|\bar{e}_{k}\rangle$ ($1\le k \le m$), $|c_{0}\rangle$ and $|c_{k}\rangle$ ($1\le k \le m$). The operations of ${x_i} \wedge {x_j}$ and ${\overline {{x_i}  \wedge {x_j}} }$ can be realized by a Toffoli gate when the target bit is initially set to 0 and 1, respectively, i.e.,  $\vert x_i \rangle \vert x_j \rangle \vert 0 \rangle  \to \vert x_i \rangle \vert x_j \rangle \vert {x_i}  \wedge {x_j} \rangle $ and $\vert x_i \rangle \vert x_j \rangle \vert 1 \rangle  \to \vert x_i \rangle \vert x_j \rangle \vert {\overline {{x_i}  \wedge {x_j}} } \rangle$. Consequently, all $|\bar{e}_{k}\rangle$ and $|c_{0}\rangle$ are initially set to be $|1\rangle$, while $|c_{k}\rangle (1\le k \le m)$ are initialized as $|0\rangle$. The quantum circuit is shown in Fig. \ref{fig2}B. Only if the final value of $c_m$ is 1, do we obtain the legal cliques.

3) {\textbf{Classifying the legal cliques.}} In order to find the largest cliques in the legal cliques, we first classify the cliques $\vert x_1 . . . x_n \rangle$ into different registers by their Hamming weights ( i.e., the number of ones that appear in its binary representation $x_1 \cdots x_n$) as sketched in Fig. \ref{fig2}C. This idea can be described by the logic flowchart in Fig. \ref{fig2}D. Auxiliary qubits $z_{i+1,j+1}$ and $z_{i+1,j}$ are employed to store the results of the formulae $\left( {{x_{i + 1}} \wedge {z_{i,j}}} \right)$ and $\left( {\overline {{x_{i+1}}}\wedge {z_{i,j}}}\right)$, respectively, which are likely implemented by Toffoli gates.  For $0\le i \le n$ and $0\le j\le i$, $|z_{i+1,j}\rangle$ and $|z_{i+1,j+1}\rangle$ are all initially prepared in state $|0\rangle$. After the loops are completed, we have successfully classified the legal cliques into $n+1$ qubits $z_{n,i}$, with $i = 0, ..., n$ according to their Hamming weight from 0 to $n$. If at least $|z_{n,i}(x) \rangle = |1 \rangle$, the legal cliques with the Hamming weight $i$ exist, else if all $2^n$ $|z_{n,i}(x) \rangle = |0 \rangle$, there is no legal cliques with the Hamming weight $i$.

4) {\textbf{Identifying the legal cliques.}} After the classification of the legal cliques by their Hamming weights, we can identify the largest ones which correspond to the biggest Hamming weight among $n+1$ registers by quantum Grover search algorithm. That is, the largest cliques are labeled by $|z_{n,i_{max}}(x)\rangle = |1 \rangle$ and $i_{max}$ is the size of the largest cliques in the graph $G$. The Grover algorithm is repeatedly applied to the data register $x$ and the qubit $z_{n,i}$ where $i$ starts from $n$ and stops at $i_{max}$. For Grover algorithm, an oracle qubit $ O$ initialized in $\vert 1 \rangle$ is introduced to perform the oracle operation, i.e., to inverse those target cliques $x$ labeled by $|z_{n,i}(x)\rangle = |1 \rangle$, which is completed by a Hadarmad gate on the oracle qubit $O$ and a CNOT gate between qubits $z_{n,i}$ and $O$. This functions as $\vert x \rangle \vert z_{n,i}(x)\rangle \vert 1 \rangle  \to \vert x \rangle  \vert z_{n,i}(x)\rangle \frac{\vert 0 \rangle  - \vert 1 \rangle}{\sqrt{2}} \to (-1)^{z_{n,i}(x)}\vert x \rangle  \vert z_{n,i}(x)\rangle \frac{\vert 0 \rangle  - \vert 1 \rangle}{\sqrt{2}}$. Thus by applying the reversal quantum circuit of the steps 2) and 3), one can accomplish the oracle operation on the data register: $\sum_{x \in \{\{ \mbox{illegal cliques} \} \cap \{\mbox{ legal cliques} |_{z_{n,i}(x) = 0} \}\}} \vert x \rangle - \sum_{x \in \{ \mbox{legal cliques} |_{z_{n,i}(x) = 1}\}} \vert x \rangle$. Then the oracle operation is followed by Hadamard transforms and a conditional phase shift on the data register to complete one iteration of Grover algorithm \cite{grover}. According to the results of Grover algorithm \cite{grover}, one requires $O(\sqrt{2^n / M})$ Grover iterations in order to obtain a solution to the search problem with high probability. Here $M$ is the number of the solutions. When $M$ is unknown, quantum counting algorithm \cite{count} can be used to approximately estimate the number $M$ of solutions.

\subsection{The example}

\begin{figure*}[t]
\centering
\includegraphics[width=16.2 cm, clip=True]{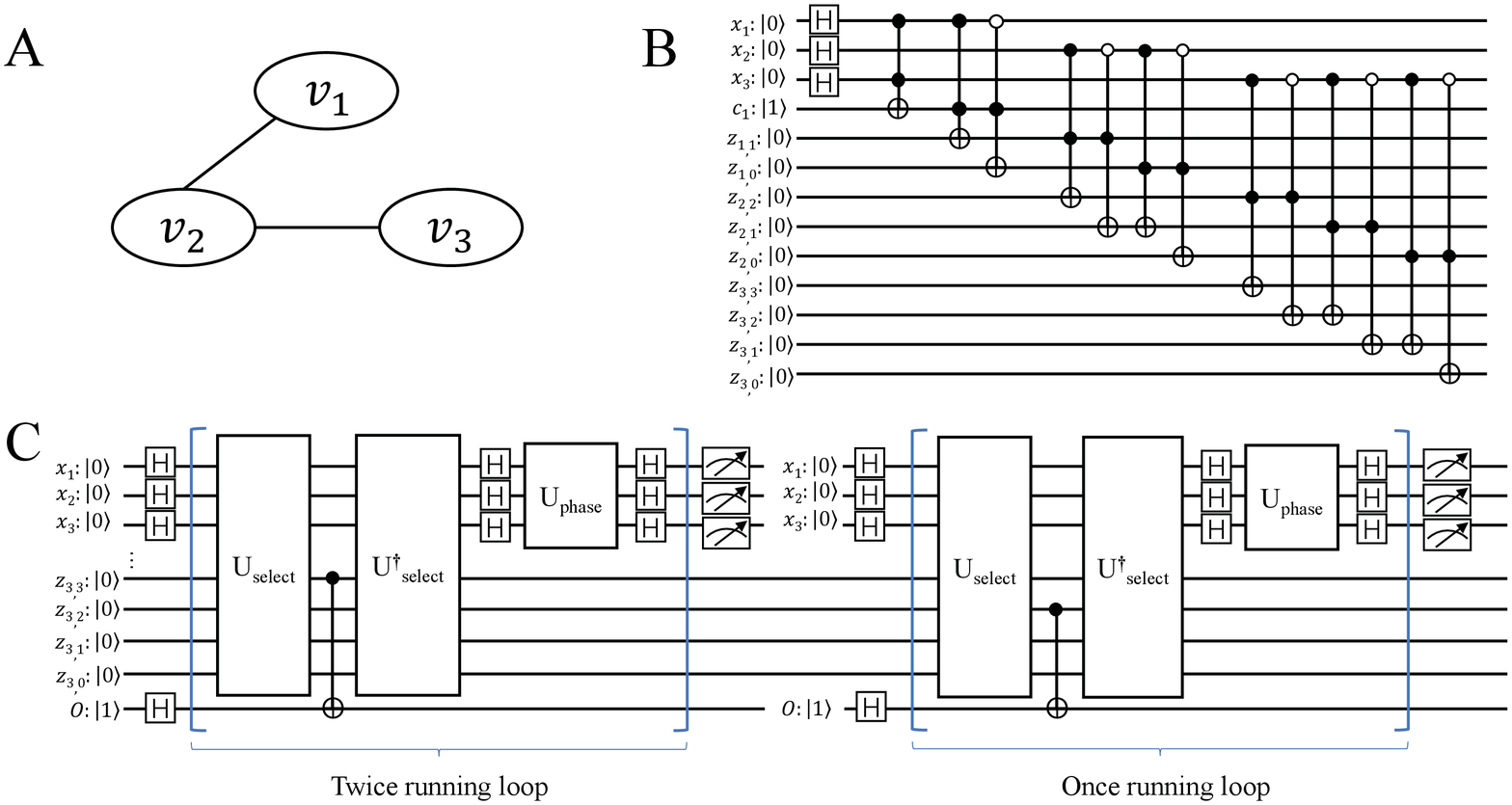}
\caption{\label{figS1} ({\bf A}) Graph $G(3,2)$ with $V = \{v_1, v_2, v_3 \}$ and $E = \{e_1 = ( v_1, v_2 ), e_2 = ( v_2, v_3 ) \}$. ({\bf B}) Quantum circuit to perform the algorithm. ({\bf C}) Diagram of quantum circuit to complete the algorithm. $U_{select}$ denotes the execution of the quantum circuit in ({\bf B}). The oracle $|O\rangle$ initialized in $\vert 1 \rangle$ is introduced for Grover iterations which include the oracle's operation, Hadamard gates and a conditional phase-shift on the data register, as explained in the text. Grover iterations are repeatedly applied to qubits from $z_{3,3}$ to $z_{3,0}$, and stop when we find a valid solution. }
\label{figS1}
\end{figure*}

To further clarify the algorithm introduced above, we present in Fig. \ref{figS1}A a simple example for a graph $G=(3,2)$ with $V = \{v_1, v_2, v_3 \}$ and $E = \{e_1 = ( v_1, v_2 ), e_2 = ( v_2, v_3 ) \}$, whose complementary graph is $\bar{G}=(3,1)$ with $ V = \{v_1, v_2, v_3 \}$ and $\bar{E} = \{\bar{e}_1 = ( v_1, v_3 )\}$. In this case, three qubits, as the data register, are initialized in the equal superposition state $\frac{1}{2 \sqrt{2}}  \sum_{x = 0}^{7} \vert x \rangle$. According to $\bar{G}$, a Toffoli gate among $v_1$, $v_3$ and $\bar{e}_1$ transforms the initial state $\frac{1}{2 \sqrt{2}}  \sum_{x = 0}^{7} \vert x \rangle \vert \bar{e}_1 = 1 \rangle$ into $ \frac{1}{2 \sqrt{2}} ( \sum_{x \in \mbox{illegal}} \vert x \rangle \vert \bar{e}_1 = 0 \rangle +\sum_{ x \in {\mbox{legal}}} \vert x \rangle \vert \bar{e}_1 = 1 \rangle  )$. Thus the illegal cliques $x = x_1 x_2 x_3 \in \{ (x_1 = 1) \& (x_3 = 1) \}$ are separated from the legal ones by the state of the qubit $\bar{e}_1$, i.e., the legal cliques are labeled by $\vert \bar{e}_1 = 1 \rangle$ in the second term. Here the register $\vert c_0 \rangle ... \vert c_m \rangle$ can be omitted because of only one edge in $\bar{G}$. Let $\vert z \rangle = \vert z_{1,0} z_{1,1} ... z_{3,0} z_{3,1} ... z_{3,3}\rangle $.  After the classification by the Hamming weight, the output state is
$ \frac{1}{2 \sqrt{2}} ( \sum_{x \in \mbox{illegal}} \vert x \rangle \vert \bar{e}_1 = 0 \rangle  \vert z  = 0 \rangle +\sum_{ x \in {\mbox{legal}}} \vert x \rangle \vert \bar{e}_1 = 1 \rangle \vert z(x) \rangle )$ where the values of $z(x)$ are listed in TABLE \ref{table1}. Thus the legal clique $x$ with the Hamming weight $i$ is classified into the qubit $z_{3,i}(x)$, i.e., $z_{3,1}$ is used to store the legal cliques with only one of $x_1,x_2,x_3$ being in 1. $z_{3,1}(x) = 1$ implies that $x$ exists in the legal cliques, and so on.

Then the final task is to find the maximal value $i_{max}$ with $z_{3,i_{max}}(x) = 1$ and the corresponding $x$. Grover search algorithm is employed to complete this task. Starting from $z_{3,3}$, due to all $z_{3,3}(x) = 0$, Grover iterations fail to return a valid solution, illustrating no legal cliques with the Hamming weight 3. Then we execute the algorithm by restarting and in the final step Grover algorithm is moved to $z_{3,2}$. With around 4 applications of Grover iteration, we may finally obtain two valid solutions $|110\rangle$ and $|011\rangle$, implying that the largest cliques are $\{v_1, v_2\}$ and $\{v_2, v_3\}$. The algorithm ends. The quantum circuit for the example is shown in Fig. \ref{figS1}B and C.

\subsection{Estimate of the complexity}

It is not difficult to access the complexity of this quantum algorithm. Given a graph with $n$ vertices, the required number of the qubits, including those as auxiliary, is at most $2m+n+2+n(n+3)/2$, implying the spatial complexity $O(n^2)$, a polynomial increment with $n$.
However, the time complexity of the Grover search algorithm is $O(\sqrt {{2^n}} )$. This implies that, in addition to the polynomial time increase of oracle's job for identifying the target states, the time complexity of our algorithm is at most $O(n^{3}\sqrt {{2^n}})\sim O(\sqrt {{2^n}})$ in the worst case. Therefore, the maximal clique problem can be solved with a quadratic speed-up by our quantum algorithm in comparison with that by the classical counterparts. Some detailed discussion can be found in the Appendix A.

\begin{figure*}[t]
\includegraphics[width=15.3 cm, clip=True]{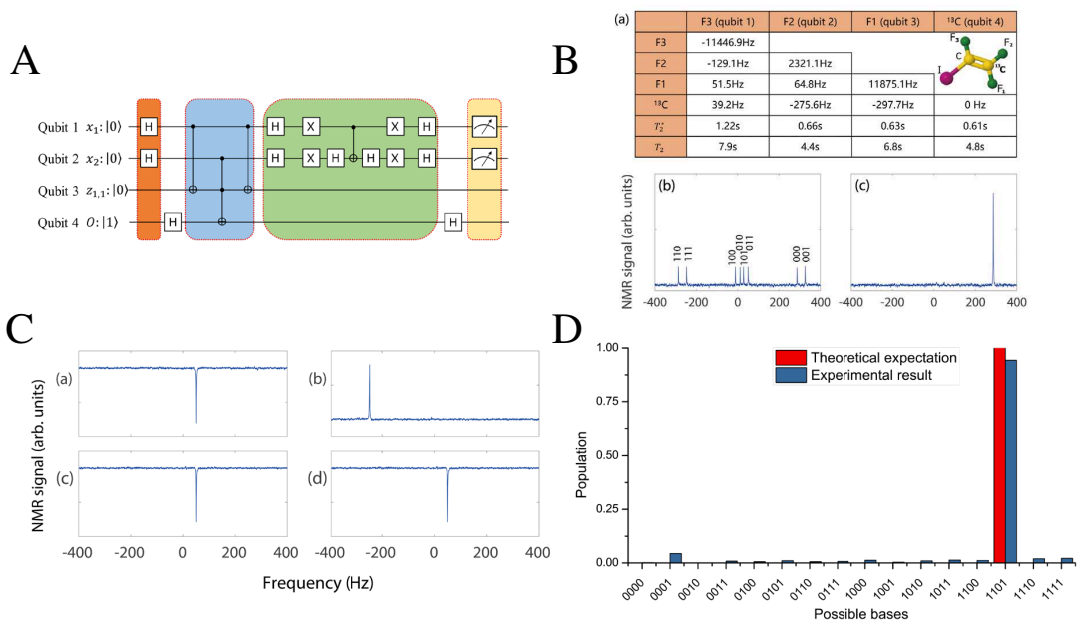}
\caption{\label{fig3} NMR experiment for a maximal clique problem of the graph $G(2,1)$ ({\bf A}) Simplified quantum circuit for solving the clique problem of the graph $G(2,1)$. ({\bf B}) Relevant parameters of the iodotrifluoroethylene molecule (a) along with the experimental $^{13}$C spectra of the thermal equilibrium state (b) and the pseudo-pure state (c) after a $\pi/2$ pulse on $^{13}$C. Eight resonance lines of $^{13}$C are labeled by the corresponding states of the three $^{19}$F spins. ({\bf C}) Experimental spectra after performing the quantum circuit ({\bf A}) by a $\pi/2$ readout pulse to spin $i$. The spectra of $^{19}$F spins were recorded via the ${}^{13}$C channel, by applying SWAP gates. (a) - (d) denote the ones for $^{13}$C, F$_1$, F$_2$, F$_3$, respectively. ({\bf D}) Reconstructed populations of the states from $|0000 \rangle$ to $|1111 \rangle$ from the experimental results in comparison with the theoretical expectations. The final state is evolved close to $|1101 \rangle$ encoding the solution of the clique problem $|x_1 x_2\rangle=|11\rangle$, i.e., $\{v_1,v_2\}$. }
\end{figure*}

\begin{table}[t]
  \centering
  \caption{Values of $z(x)$ in the output state after the quantum circuit in Fig. \ref{figS1}B is performed.}\label{output}
  \begin{tabular}{|ccc|c|cc|ccc|cccc|}
\hline
 $x_1$ & $x_2$ & $x_3$ &  $\bar{e}_1$ & $z_{1,1}$ & $z_{1,0}$ & $z_{2,2}$ & $z_{2,1}$ & $z_{2,0}$ & $z_{3,3}$ & $z_{3,2}$ & $z_{3,1}$ & $z_{3,0}$\\
\hline
  0    &  0    &   0   &  1     &   0       &      1    &   0       &      0    &   1       &     0     &       0   &      0    &   1 \\
  0    &  0    &   1   &  1     &   0       &      1    &   0       &      0    &   1       &     0     &       0   &      1    &   0 \\
  0    &  1    &   0   &  1     &   0       &      1    &   0       &      1    &   0       &     0     &       0   &      1    &   0 \\
  0    &  1    &   1   &  1     &   0       &      1    &   0       &      1    &   0       &     0     &       1   &      0    &   0 \\
  1    &  0    &   0   &  1     &   1       &      0    &   0       &      1    &   0       &     0     &       0   &      1    &   0 \\
  1    &  0    &   1   &  0     &   0       &      0    &   0       &      0    &   0       &     0     &       0   &      0    &   0 \\
  1    &  1    &   0   &  1     &   1       &      0    &   1       &      0    &   0       &     0     &       1   &      0    &   0 \\
  1    &  1    &   1   &  0     &   0       &      0    &   0       &      0    &   0       &     0     &       0   &      0    &   0 \\
\hline
\end{tabular}
\label{table1}
\end{table}

\section{experimental implementation}

To verify our proposed algorithm, we have also accomplished a proof-of-principle NMR experiment for the simplest clique problem for a graph $G=(2,1)$ that consists of two vertices $\{v_2,v_1\}$ and an edge $\{(v_1,v_2)\}$. After optimizing the quantum circuit, as detailed in Appendix B, we only require four qubits for solving this problem, as shown in Fig. \ref{fig3}A. The four qubits $|x_1 \rangle$, $|x_2 \rangle$, $|z_{1,1} \rangle$ and $|O\rangle$ are initially prepared to $|0 \rangle$, $|0 \rangle$, $|0 \rangle$, $|1 \rangle$, respectively.
The experiment is carried out on a Bruker AV-$400$ NMR spectrometer (9.4 T) at $303.0$ K with the sample iodotrifiuoroethylene C$_2$F$_3$I dissolved in $d$-chloroform, where three ${}^{{\rm{19}}}$F nuclei and a ${}^{{\rm{13}}}$C nucleus constitute a four-qubit quantum processor. The natural Hamiltonian of this four-spin quantum system in the double-rotating frame is given by \cite{NMR quantum computation, spin dynamics},
\begin{equation}
\label{hamiltonian}
{H_\text{NMR}} = \sum\limits_{j = 1}^4 {\pi {\nu_j}\sigma _z^j + \sum\limits_{1 \le j \le k \le 4} {\frac{\pi }{2}} } {J_{jk}}\sigma _z^j\sigma _z^k,
\end{equation}
where the measured parameters are shown in Fig. \ref{fig3}B.  The chemical shifts ${{\nu_j}}$ and the $J$-coupling constants $J_{jk}$ are, respectively, listed in the diagonal and off-diagonal terms.

We first prepared a pseudo-pure state $\rho_{0000}$ using the line-selective pulse method \cite{line-selective pulses} with the fidelity of $97.23\%$. Here the fidelity is calculated by $F=\sqrt{\langle 0000 | \rho_{{\rm{0000}}} | 0000 \rangle}$. Then we used a high-fidelity shaped pulse found by the gradient ascent pulse engineering (GRAPE) algorithm \cite{grape} to realize the quantum circuit in Fig. \ref{fig3}A.  The GRAPE pulse has a duration of 26 ms, with a theoretical fidelity above 99.85$\%$. The last step is to measure the output state encoding the solution of the clique problem, which only requires the occupation information on the computational basis states from $\vert 0000 \rangle$ to $\vert 1111 \rangle$. To reconstruct the populations, we recorded the experimental spectrum for each spin after a $\pi/2$ readout pulse to this spin.  Since the natural abundance of ${}^{13}$C in the sample is about $1\%$, we read out all four spins via the ${}^{13}$C channel, by applying SWAP gates between each $^{19}$F spin and the $^{13}$C spin, to distinguish those molecules against the large background.
The experimental spectra are shown in Fig. \ref{fig3}C, where the intensities of the main resonant lines are respectively $-0.9085, 0.9122, -0.9381$ and $-0.9672$ in comparison to the pseudo-pure state. From these, the reconstructed populations \cite {tomo} on the 16 computational basis states are plotted in Fig. \ref{fig3}D where  the population on $|1101\rangle$ is around 0.9429, much larger than those of other states in Fig. \ref{fig3}D. Therefore this experimental result implies that we found the solution of the clique problem of the graph $G(2,1)$: $|x_1 x_2\rangle=|11\rangle$, i.e., $\{v_1,v_2\}$, with a high probability 94.29\%.

The experimental errors are mainly caused from the imperfect initial state preparation ($\sim 2.8\%$), the GRAPE pulse error ($\sim 1.0\%$) and the imperfect readout pulses ($\sim 2.0\%$). Decoherence during the implementation is negligible due to the fact that the experimental running time is less than 30 ms, much shorter than the shortest relaxation time 600 ms.

\section{conclusion}

In summary, we have proposed an optimal oracle-related quantum algorithm based on a quantum circuit oracle model to solve the maximal clique problem for any graph $G$ with $n$ vertices and $\theta$ edges. Our NMR experimental performance, although only solving a simplest clique problem, gives us hopes that if our quantum algorithm really works efficiently, we would have reason to eagerly await the quantum computer capable of running larger numbers of qubits to practically treating the graph-relevant NP-complete problems with quadratic speed-up.

An open question regarding the extension of our quantum algorithm is whether other NP-complete problems can also be optimally solved if they are reduced to the clique problem. Our current answer is negative since solving a certain NP-complete problem by this way would probably take a time complexity more than $O(2^{n})$ (See Appendix C for details), which makes the quantum treatment less efficient than the classical treatment (with time complexity of $O(2^{n})$). Further clarification for this point is underway.

\section*{ACKNOWLEDGMENTS}
This work is supported by National Key Basic Research Program of China (2013CB921800, 2014CB848700, 2017YFA0304503), National Science Fund for Distinguished Young Scholars of China (Grant No. 11425523), National Natural Science Foundation of China (Grants No. 11674360, No. 11375167 and No. 11227901), the Strategic Priority Research Program (B) of the CAS (Grants No. XDB01030400 and No. XDB21010100), Key Research Program of Frontier Sciences of the CAS (Grant No. QYZDY-SSW-SLH004).

\begin{widetext}
\appendix
\section{Complexity Assessment}
To solve the maximal clique problem of a graph $G$ with $n$ vertices and $\theta$ edges (the complementary graph $\bar G$ thus has $n$ vertexes and $m=\frac{n(n-1)}{2}-\theta$ edges), we require $n$ Hadamard gates to generate the uniform superposition state, $2m$ Toffoli gates to exclude illegal cliques, and $n(n+1)$ NOT gates and $n(n+1)$ Toffoli gates to classify legal cliques by their Hamming weights. This indicates that we also need $n(n+1)$ NOT gates and $2m+n(n+1)$ Toffoli gates to restore the qubits to their initial states. The steps described above, including a CNOT and a Hadamard gate, are employed to complete the oracle's work of the Grover search. Then Grover search is performed. The Grover's operator can be decomposed into $H^{\otimes n}U_{\text{PSG}}H^{\otimes n}$, where $U_{\text{PSG}}$ is an $n$-qubit conditional phase-shift gate defined as
\begin{equation}
U_{\text{PSG}}:\begin{cases}|x\rangle\rightarrow -|x\rangle & x\neq0\\
            |0\rangle\rightarrow|0\rangle.
            \end{cases}
\end{equation}
Thus one requires $O(\sqrt{2^n / M})$ Grover iterations in order to obtain a solution to the search problem with high probability. Here $M$ is the number of the solutions.
Therefore the numbers of logic gates required to solve the clique problem are as follows,
\begin{equation}
\begin{cases}
\text{Hadamard gate}&O(2n2^{\frac{n}{2}}+n+1)\sim O(n2^{\frac{n}{2}})\\
\text{NOT gate}&O(2(n^2+n)2^{\frac{n}{2}})\sim O(n^22^{\frac{n}{2}})\\
\text{CNOT gate}& O(2^{\frac{n}{2}})\\
\text{Toffoli gate}& O((4m+2(n^2+n))2^{\frac{n}{2}})\sim O((m+n^2)2^{\frac{n}{2}})\\
\text{Conditional phase-shift gate of $n$ qubits} &O(2^{\frac{n}{2}})\\
\text{Measurement}&O(1).
\end{cases}
\end{equation}
For the worst case, we need to repeat the algorithm $n$ times. Thus the complexity is
\begin{equation}
\begin{cases}
\text{Hadamard gate}&O(n(2n2^{\frac{n}{2}}+n+1))\sim O(n^22^{\frac{n}{2}})\\
\text{NOT gate}&O(2n(n^2+n)2^{\frac{n}{2}})\sim O(n^32^{\frac{n}{2}})\\
\text{CNOT gate}& O(n2^{\frac{n}{2}})\\
\text{Toffoli gate}& O(n(4m+2(n^2+n))2^{\frac{n}{2}})\sim O(n(m+n^2)2^{\frac{n}{2}})\\
\text{Conditional phase shift gate of $n$ qubits} &O(n2^{\frac{n}{2}})\\
\text{Measurement}&O(n).
\end{cases}
\end{equation}
As such, the average gate complexity is equal to $(1+2+\cdots+n)/n$ multiplied by the complexity of the best case,
\begin{equation}
\begin{cases}
\text{Hadamard gate}&O(\frac{n+1}{2}(2n2^{\frac{n}{2}}+n+1))\sim O(n^22^{\frac{n}{2}})\\
\text{NOT gate}&O((n+1)(n^2+n)2^{\frac{n}{2}})\sim O(n^32^{\frac{n}{2}})\\
\text{CNOT gate}& O(\frac{n+1}{2}2^{\frac{n}{2}})\sim O(n2^{\frac{n}{2}})\\
\text{Toffoli gate}& O((n+1)(2m+(n^2+n))2^{\frac{n}{2}})\sim O(n(m+n^2)2^{\frac{n}{2}})\\
\text{Conditional phase shift gate of $n$ qubits} &O(\frac{n+1}{2}2^{\frac{n}{2}})\sim O(n2^{\frac{n}{2}})\\
\text{Measurement}&O(\frac{n+1}{2})\sim O(n).
\end{cases}
\end{equation}
In experiments the Hadamard gates for initializing the system can be implemented in parallel, which means the time complexity associated with Hadamard gates for initialization should actually be divided by $n$. Nevertheless, this does not affect the asymptotic complexity since its dominant part is due to the Hadamard gates used in the Grover's operator.

The spatial complexity can be obtained by counting the number of qubits required in the algorithm. In summary, one needs $n$ data qubits to encode the graph with $n$ vertices, $2m+1$ auxiliary qubits ($\bar{e}$ and $c$ quantum registers) to exclude illegal cliques, and $\frac{n(n+3)}{2}$ qubits ($z$ quantum register) to classify legal cliques. In addition, the qubit $|O\rangle$ is needed for the Grover search. So totally, $O(2m+n+2+\frac{n(n+3)}{2})\sim O(n^2+m)$ qubits are required. The numbers of the qubits needed in both the best and the worst cases are the same since qubits can be reused. Note that "$\sim$" means asymptotically equivalent ($g(n)$ is said to be asymptotically equivalent to $f(n)$ if $\lim_{n\to \infty}g(n)/f(n)<\infty$).

Therefore, we consider that our proposed oracle-related quantum algorithm for the maximal clique problem behaves with polynomial-scaled spatial complexity and $O(\sqrt{2^{n}})$-scaled time complexity. Based on the fact that any oracle-related quantum algorithm cannot work better than quadratic speed-up over its classical counterparts \cite{NP1,NP2}, we identify our proposed quantum algorithm to be optimal with respect to currently known quantum algorithms.

\section{Quantum circuit for our experiment}

This section is to explain how to reach a four-qubit quantum circuit, as plotted in Fig. 4A, for a clique problem regarding the graph $G$ with two vertices and one edge.

\begin{figure}[t]
\centering
\includegraphics[width=16.2 cm, clip=True]{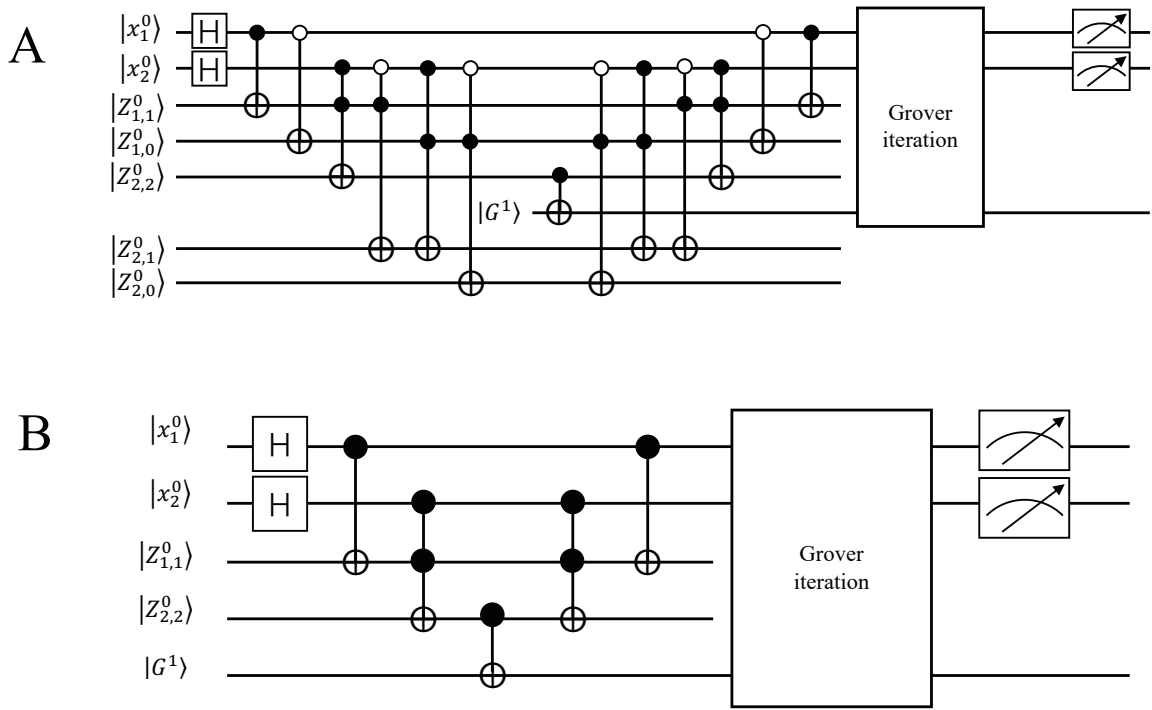}
\caption{\label{figS2} (A) The standard version of the quantum circuit for solving the clique problem; (B) Simplified quantum circuit for solving the clique problem. The superscript in $|\cdot\rangle$ represents the initial state of the qubit in $|0\rangle$ or $|1\rangle$.}
\label{figS2}
\end{figure}

All the possible cliques in $G$ are $\{v_1,v_2\}$, $\{v_2\}$, $\{v_1\}$ and $\varnothing$, where the maximum-sized clique is $\{v_1,v_2\}$. The quantum circuit based on the standard steps as described in the main text should involve seven qubits, see Fig. \ref{figS2}A. However, since there is no clique in the complementary graph of $G$, we can skip the step of excluding  illegal cliques and simplify the circuit. According to the flowchart in Fig. 2A, we describe the algorithm in Fig. \ref{figS2}B, where the qubits $|z_{2,1}^0 \rangle$ and $|z_{2,0}^0 \rangle$ are excluded from the circuit, because we can obtain the answer in the first Grover iteration and measurement. Besides, the qubit $|z_{1,0}^0 \rangle$ works as a control on the qubits $|z_{2,1}^0 \rangle$ and $|z_{2,0}^0 \rangle$. Since we have deleted  $|z_{2,1}^0\rangle$ and $|z_{2,0}^0 \rangle$, the qubit $|z_{1,0}^0 \rangle$ can also be discarded.

In fact, the quantum circuit can be further simplified. Since the qubit $|z_{2,2}^0 \rangle$ is designed only to control the oracle qubit $|O \rangle$, this qubit $|z_{2,2}^0 \rangle$ can also be reduced if we carefully consider the initial state of $|O \rangle$. Therefore we finally get to the four-qubit quantum circuit, as in Fig. 4A, for solving the clique problem regarding the graph $G$.

\section{Extension of our quantum algorithm}

We argue below that it is impossible to optimally solve other NP-complete problems by simply reducing them to the clique problem under consideration.

For simplicity, we exemplify the three-satisfiability (3-SAT) problem with $n$ Boolean variables and $m$ clauses, in which each clause contains three Boolean variables. The solution to 3-SAT is to find whether there is a truth assignment that satisfies all the clauses. In terms of the approaches in \cite{1,2,3}, such a 3-SAT problem can be reduced to a clique problem with $(2n + 3m)$ vertices and $[(2n+3m)(2n+3m-1)/2-(n+6m)]$ edges. But solving such a clique problem, with our optimal oracle-related quantum algorithm, will take a time complexity $O(\sqrt{2^{2n+3m}})> O(2^{n})$. This implies that, by this reduction way, the quantum solution of the 3-SAT problem is less efficient than a classical treatment.

Further exploration is needed to clarify this problem. If this reduction way is in principle unavailable for quantum treatment, developing independent quantum algorithms for different NP-complete problems will be indispensable.

\end{widetext}

\end{document}